# Synthesis, Engineering, and Theory of 2D van der Waals Magnets

M. Blei[1], J.L. Lado[2], Q. Song[3], D. Dey[4], O. Erten[4], V. Pardo[5,6], R. Comin[3], S. Tongay[1], A.S. Botana[4]

[1] Materials Science and Engineering, School for Engineering of Matter, Transport and Energy, Arizona State University, Tempe, Arizona 85287, USA
[2] Department of Applied Physics, Aalto University, 00076 Aalto, Espoo, Finland
[3] Department of Physics, Massachusetts Institute of Technology, Cambridge, MA 02139, USA
[4] Department of Physics, Arizona State University, Tempe, Arizona, 85287, USA
[5] Departamento de Fisica Aplicada, Universidade de Santiago de Compostela, E-15782 Campus Sur s/n, Santiago de Compostela, Spain
[6] Instituto de Investigacions Tecnoloxicas, Universidade de Santiago de Compostela, E-15782 Campus Sur s/n, Santiago de Compostela, Spain

## 1. INTRODUCTION

The recent discovery of magnetism in monolayers of two-dimensional (2D) van der Waals (vdW) materials has opened new venues in Materials Science and Condensed Matter Physics. Until recently, 2D magnetism remained elusive since the existence of magnetic order in 2D is a priori not guaranteed. The story changed in 2016 when two groups provided evidence of antiferromagnetism in monolayers of $FePS_3$ [1,2]. In 2017, the presence of ferromagnetic order was proven in monolayers of $CrI_3$ and on a bilayer of $Cr_2Ge_2Te_6$ [3,4]. The list of candidates has been growing ever since [5-7].

There are various aspects that make 2D vdW crystals with magnetic order very interesting. First, magnetic order in 2D can only happen if there is no continuous rotational symmetry, otherwise the proliferation of low-energy spin waves, that lies behind the Mermin-Wagner theorem [8], destroys magnetic order at any finite temperature. Therefore, magnetic anisotropy and spin waves control the transition temperature of 2D magnets and play a much more important role than in their 3D counterparts. Second, the electronic and mechanical properties of 2D materials can be widely tuned in various ways: by gating, proximity, and chemical functionalization, which permits to conceive devices where magnetic order is controlled at will. Third, magnetic order adds a new functionality to the set of Lego-like pieces that enriches the game of vertical integration of 2D materials in van der Waals heterostructures. The stacking of materials with magnetic order, superconducting order, spin-valley coupling, and graphene will probably result in structures with completely new and unexpected properties that we can now explore both theoretically and experimentally.

Within just three years, the discovery of 2D magnets has already opened up new opportunities in spintronics i.e. spin pumping devices, spin transfer torque, and tunneling magnetoresistance [9-11]. We envision future applications that may extend into other realms, including sensing and data storage. Here, we review some of the experiments in which magnetism in strictly 2D has been confirmed. We discuss common synthesis techniques for these materials and methods for engineering their magnetic properties. Further, we analyze in detail some of the most important theoretical aspects that need to be considered to understand 2D magnets. Finally, we identify different phenomena that we anticipate will be the next steps to follow in the field.

## 2. EXPERIMENTAL FINDINGS ON 2D MAGNETS TO DATE

### 2.1 Evidence of magnetic order in 2D

We first review some of the existing experiments providing evidence for magnetic order at the monolayer level (or close to it) and describe briefly the corresponding 2D vdW materials (see Table I).

***Antiferromagnetism in FePS$_3$.*** The first experimental evidence of magnetic order at finite temperature in monolayers was found in FePS$_3$ in 2016. By monitoring the Raman peaks that arise from zone folding due to antiferromagnetic ordering (Fig. 1a), it was demonstrated that FePS$_3$ exhibits antiferromagnetic ordering down to the monolayer limit with a $T_N$ as high at 118 K [1,2].

***Ferromagnetism in CrI$_3$.*** Kerr microscopy experiments have shown that ferromagnetism in this material persists down to the monolayer level (Fig. 1b) with a large critical temperature of 45 K (not far from that of the bulk ~ 61 K) [3]. Magnetic order in this compound shows an out-of-plane easy axis anisotropy. The bilayer system (Fig. 1c) showed a surprising lack of Kerr signal attributed to an interlayer antiferromagnetic arrangement genuine of the bilayer. Ever since, there has been intense interest in trying to elucidate the importance of stacking order for the magnetic response of this material [12,13].

***Antiferromagnetism in CrCl$_3$.*** Tunneling magnetoresistance measurements in few-layer CrCl$_3$ provided early evidence of antiferromagnetic ordering down to bilayer samples, shown in Fig. 1d. Few-layer samples preserved the same magnetic ordering as their bulk counterparts, with in-plane easy-axis anisotropy, and antiparallel spin ordering between layers. Strikingly, ultrathin CrCl$_3$ samples showed a tenfold increase in exchange energy, which was attributed to the different stacking order and its feedback on the out-of-plane exchange interactions at low temperatures [14].

***Ferromagnetism in Cr$_2$Ge$_2$Te$_6$.*** Magnetic order at the bilayer level was probed in Cr$_2$Ge$_2$Te$_6$ by means of Kerr rotation experiments [4]. The monolayer, in turn, was found to degrade rapidly. The magnetic transition temperature proved to be tunable by means of an external magnetic field. This clearly shows the potential to build devices based on 2D vdW magnets with properties that can be easily manipulated.

***Ferromagnetism in Fe$_3$GeTe$_2$ (FGT).*** Itinerant ferromagnetism persists in Fe$_3$GeTe$_2$ down to the monolayer limit with a sizable out-of-plane magnetocrystalline anisotropy [15]. Magnetism was studied by probing the Hall resistance, shown in Fig. 1e. The ferromagnetic transition temperature is suppressed relative to the bulk (205 K) but an ionic gate can raise $T_c$ all the way up to room temperature, opening up opportunities for potential voltage-controlled magnetoelectronics[16].

***Ferromagnetism in transition-metal dichalcogenides (TMDs).*** A strong ferromagnetic signal at room temperature has been reported at the single-layer level in VSe$_2$. However, spontaneous ferromagnetism in this system remains a controversial issue due to the possibility of charge density wave (CDW) formation and the subsequent suppression of magnetic order [17,18]. Angle resolved photoemission spectroscopy (ARPES) and scanning tunneling microscopy (STM) have revealed an electronic reconstruction of single layer VSe$_2$ without a detectable FM exchange splitting, casting doubts on whether magnetism originates from an induced band structure spin splitting, or if extrinsic defects come into play [19-21]. Room temperature ferromagnetism has also been reported in MnSe$_2$ films grown by MBE. From SQUID measurements in the monolayer limit, the magnetic signal is assigned to intrinsic ferromagnetism with a $T_c$ close to room temperature [22].

| Material | Magnetic order | $T_c$ | Magnetic lattice | Refs. |
|---|---|---|---|---|
| $FePS_3$ | AFM | zig-zag? | honeycomb | 1, 2 |
| $CrI_3$ | FM | 45 K | honeycomb | 3 |
| $CrCl_3$ | AFM | 14 K | honeycomb | 14 |
| $Cr_2Ge_2Te_6$ | FM | 45 K | honeycomb | 4 |
| $Fe_3GeTe_2$ | FM | 300 K | triangular | 15 |
| $VSe_2$ | FM | 300 K | triangular | 17 |
| $MnSe_2$ | FM | 300 K | triangular | 22 |

**Table I.** vdW material systems for which long range magnetic order has been confirmed experimentally in 2D and their characteristics. AFM stands for antiferromagnetic and FM for ferromagnetic.

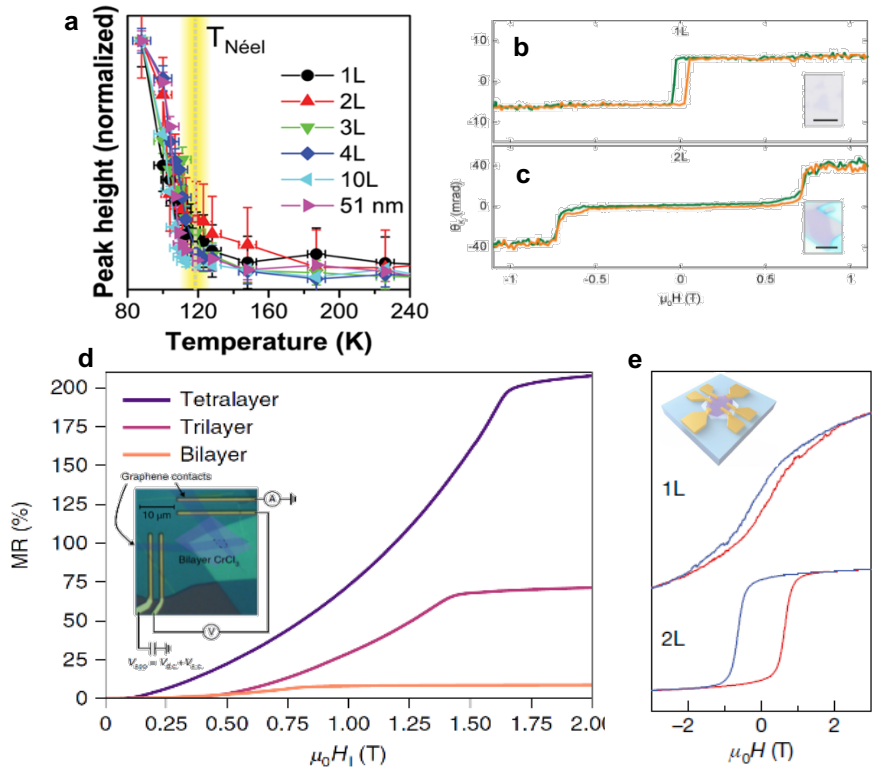

**Figure 1.** Characterization of magnetism in 2D. (a) Temperature dependence of the characteristic Raman peak resulting from spin order-induced folding of the Brillouin zone in $FePS_3$ as a function of thickness[2]. (b,c) Magneto-optical Kerr rotation as a function of applied magnetic field in flakes of $CrI_3$, revealing a ferromagnetic (antiferromagnetic) ground state in monolayer (bilayer) samples. The insets show optical microscope images of $CrI_3$ (scale bars 5 μm)[3] (d) Magnetoresistance versus in-plane magnetic field for bilayer, trilayer and tetralayer tunnel junctions at 4 K. The inset shows the optical microscope image of a bilayer $CrCl_3$ tunnel junction device.[14] (e) Low-temperature Hall resistance $R_{xy}$ in FGT thin-flake samples and ferromagnetic hysteresis down to the monolayer limit. Inset: schematic of the Hall effect measurement on FGT flakes.[15]

## 2.2 Common techniques for the synthesis of van der Waals magnetic crystals.

After discussing the early magnetic measurements and observations done on 2D magnetic crystals, we focus on common synthesis techniques used for producing layered vdW magnetic crystals. At the time of writing this review article, there are no studies that enable researchers to produce monolayer or few-layer thick magnetic crystals at large scales using commercially compatible chemical vapor deposition (CVD) or atomic layer deposition (ALD) methods. This is mainly because of the limited environmental stability of some of these 2D magnetic crystals and/or lack of established surface chemistry routes to enable layer-by-layer deposition.

Because of these limitations, the community currently heavily relies on the production of high crystalline quality, defect free crystals that are ideally free of magnetic impurities such as Fe, Co, and Ni. Once these layered crystals are produced, a routine mechanical exfoliation [23] technique is used to isolate monolayer to few-layer thick sheets onto the desired substrates. Depending on the crystal type (halide vs. chalcogen) as well as on their phase diagrams, different crystal growth techniques are used to produce these materials. Based on the most common studies vdW magnetic crystals in the field, the popular growth techniques include chemical vapor transport (CVT), sublimation, or flux zone techniques.

*Chemical vapor transport.* The first growth technique, namely chemical vapor transport (CVT), is an extremely effective and reliable method used to produce macroscale layered materials, including transition metal thiophosphates, FGT, and TMDs discussed in this article [24-26]. CVT involves transporting precursors from hot to cold zone for endothermic, or cold to hot zone for exothermic reactions using transport agents [27]. To produce crystals using CVT, it is necessary to evacuate and seal stoichiometric amounts of precursors inside thick (usually 1mm-2mm) quartz ampoules before carrying out well engineered thermal processing (crystal growth). The growth temperature (cold and hot zone) is carefully selected based on the binary or ternary phase diagrams, and usually involves high temperature processes that naturally build very high vapor pressures. For these reasons, thicker wall quartz ampoules are required for the growth process. In addition to the precursors, transport agents such as iodine and bromine are used as promoters for fast chemical reactions and larger crystal growth. For example, transition metal thiophosphate crystal synthesis (such as $MnPS_3$, $FePS_3$, and $CoPS_3$) typically involves the use of halides ($I_2$, $Br_2$, etc.) as transport agents to produce these vdW crystals [24]. Since $CrI_3$ crystals already contain iodine in the crystal matrix, the reaction of elemental Cr and $I_2$ in a quartz ampoule is sufficient [28].

*Sublimation.* In addition to CVT, physical vapor transport (PVT or sublimation), is another method that can be used to produce high quality, single-crystals of transition metal halides [29, 30]. It has proven to be a useful and low-cost technique that does not require sophisticated ampoule sealing processes necessary for chemical vapor transport reactions, although evacuated and sealed ampoules may also be used to carry out the growth [31]. For the synthesis of $CrCl_3$ and other transition metal halides, commercially available compounds are placed in an open-ended or sealed quartz tube and positioned in a horizontal furnace with the desired temperature gradient. No additional transport agent is required as these materials contain the necessary halide, and self-transport. After heating the compound to their respective sublimation points, the transition metal halides transport to the cold zone of the furnace, where they nucleate and grow directly on the walls of the quartz tube. Large, high-quality crystals can be obtained in 24-48 hrs [32].

***Flux zone growth.*** While the previously discussed vapor-phase synthesis techniques can be used for transition metal halides, other classes of 2D vdW magnets, such as $Cr_2Ge_2Te_3$ and $Fe_3GeTe_2$ can be obtained from solution-phase flux methods to achieve large, high-quality crystals [33-35]. For this method, stoichiometric amounts of precursors and flux (solvent) are loaded into an inert crucible, such as quartz, and vacuum sealed (~$10^{-5}$ torr). In the case of $Cr_2Ge_2Te_6$, excess germanium and tellurium are used to create a self-flux. As this is a solution-based technique, where the precursors and flux are in direct contact with the crucible, careful selection of the latter must be made to ensure that no undesirable reactions occur between the crucible and the precursors during the growth. Additionally, the inorganic flux is chosen, among other factors, to have a high solubility of the desired elements at the growth temperature, not to create any competing phases, and in many cases it can be part of the chemical composition of the resultant product.

The precursors and flux are then heated above their melting temperatures and slowly cooled over a period of several days. Growth parameters are selected based on prior knowledge of the product's binary or ternary phase diagrams, but experimentally determining and optimizing the growth parameters is often necessary because phase diagrams have not been established for the desired material system. After the growth is complete, the final step is to remove the flux from the crystal. The most common method to remove the flux from tellurium-based vdW magnets is to melt the tellurium and remove it through a centrifugation process, and depending on the flux, additional steps may be required to fully remove the flux [36].

## 2.3 Engineering magnetism in 2D

A significant advantage of 2D materials is that their physical properties are highly tunable by means of external control parameters that include electrostatic doping, pressure, and strain. Here we highlight a few recent demonstrations of tunable magnetism in 2D materials.

***Electrostatic Doping.*** Electrostatic doping is a powerful technique for tuning the electronic properties of 2D materials. The working principle is similar to that underlying field-effect transistors and is based on the direct transfer of electronic or ionic charges from a dielectric into the target 2D material. Electrostatic doping has a series of advantages over chemical doping of bulk materials. It is continuously controllable through a gate bias, and it is compatible with a variety of dopant species – from simple electrons/holes to specific ions/chemical functional groups – and it can be applied to most 2D materials without being hindered by phase separation issues in non-stoichiometric bulk synthesis. It has been shown that the electrostatic doping in 2D materials could unveil new physics, such as unconventional superconductivity in $MoS_2$[37], twisted bilayer graphene[38], or structural transitions in $MoTe_2$[39].

As mentioned above, bilayer $CrI_3$ is a layered antiferromagnet and it was found that the interlayer exchange coupling is tunable by electrostatic doping. Fig. 2a shows the schematic of the representative bilayer $CrI_3$ device. This device is a vertical stack of a bilayer $CrI_3$ flake and a graphite contact encapsulated between two hexagonal boron nitride (hBN) flakes and a graphite top gate. By applying the gate voltage to the insulator hBN, the electric dipoles at the hBN-$CrI_3$ interface introduce carrier injection into the bilayer $CrI_3$. Fig. 2b shows the reflection magnetic circular dichroism (RMCD) signal as a function of both top- and back-gate voltages near the metamagnetic transition field $\mu_0H = 0.78$ T. The red region on the right is the signal from the ↑ ↑ state, and the pink region on the left is from the layered AFM state. The dashed line boundary indicates that the metamagnetic transition could be effectively tuned by electrostatic doping[40, 41].

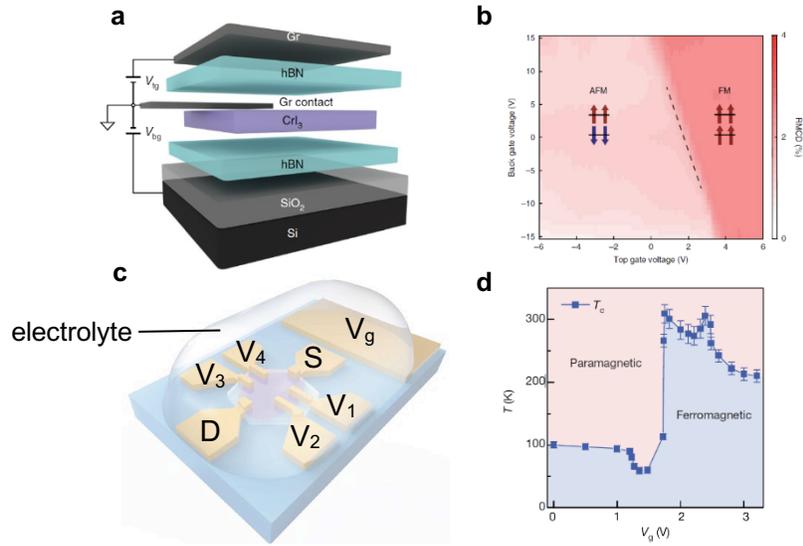

**Figure 2.** Gate-tunable magnetism of $CrI_3$ and FGT (a) Schematic of a dual-gated bilayer $CrI_3$ device. (b) RMCD signal of the electrostatic doped bilayer $CrI_3$. The dashed contour shows the tunable metamagnetic transition field through bias voltage[41]. (c) Schematic of the FGT device structure and measurement setup. S and D label the source and drain electrodes, respectively, and $V_1$, $V_2$, $V_3$ and $V_4$ label the voltage probes. The solid electrolyte ($LiClO_4$ dissolved in polyethylene oxide matrix) covers both the FGT flake and the side gate. (d) Phase diagram of the trilayer FGT sample as the gate voltage and temperature are varied. The transition temperature is determined from the extrapolation of the temperature-dependent anomalous Hall resistance to zero[15].

FGT is a layered ferromagnet with bulk $T_c$ = 205 K. In monolayer FGT, however, magnetic ordering is suppressed due to thermal fluctuations of long-wavelength acoustic-like magnon modes in 2D, while the trilayer sample has a $T_c$ ~ 100 K. Strikingly, ionic liquid gating could raise $T_c$ to room temperature, much higher than the bulk $T_c$ (Fig. 2d). This ionic gating method (Fig. 2c) intercalates the $Li^+$ from $LiClO_4$ (the transparent liquid electrolyte) onto the surface of the FGT by the gate voltage $V_g$, and the doping level could reach values as high as $10^{14} cm^{-2}$, which is one order of magnitude higher than those achievable using hBN gates[15].

***Pressure.*** In a vdW material, a small change of the interlayer spacing can cause a drastic change in physical properties. In particular, if the material supports magnetism, the interlayer interactions can be modified to produce a change in the magnitude and sign of the exchange coupling. Hydrostatic pressure is a typical method for continuous control of interlayer coupling via interlayer spacing in vdW crystals[42].

Fig. 3a shows a schematic of the experimental set-up of the high-pressure study of $CrI_3$. A magnetic tunnel junction (MTJ) device was composed of bilayer $CrI_3$ sandwiched between top and bottom multilayer graphene contacts. The entire MTJ was encapsulated by hBN to prevent sample degradation. The device was then held in a piston cylinder cell filled with oil for application of hydrostatic pressure. Magnetic states were probed by using tunneling magnetoresistance measurements as shown in Fig. 3b. After removal from the cell, reflective magnetic circular dichroism (RMCD) microscopy (Fig. 3c and 3d) showed that the bilayer $CrI_3$ irreversibly transitioned from antiferromagnetic to ferromagnetic ordering[43, 44].

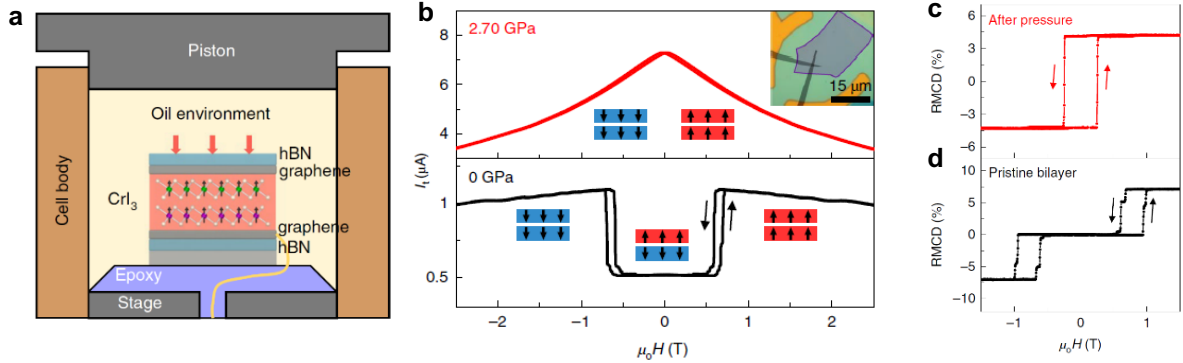

**Figure 3**. High pressure study of $CrI_3$. (a) Schematic of a high-pressure experimental set-up. The force applied to the piston exerts pressure on the bilayer $CrI_3$ device through oil. (b) Tunneling current vs. magnetic field H at two pressures. Insets: magnetic states and optical microscopy image of a bilayer device. (c, d) RMCD signal from the bilayer after removal from pressure cell where it was subjected to comparable pressure (2.45 GPa), and from a pristine bilayer $CrI_3$[44].

***Strain.*** 2D materials possess outstanding mechanical properties and can sustain larger strain than their bulk counterparts. Monolayer $MoS_2$ is predicted to sustain elastic strain levels up to 11%, and monolayer FeSe up to 6%[45, 46]. Strain engineering has been shown to be an effective approach to tune the properties of 2D materials using various methods including substrate-lattice mismatch[47, 48], mechanically actuated strain cells[49, 50], and nanomechanical drumheads[51].

Fig. 4a shows the experimental set-up used in the biaxial strain study of bilayer $CrI_3$ through the nanomechanical drumhead. By applying a voltage to the silicon substrate, the electrostatic force between silicon and $CrI_3$ can apply tensile biaxial strain to $CrI_3$ itself. Fig. 4b shows the structure of the bilayer $CrI_3$ device, with $CrI_3$ encapsulated within two stable 2D materials, few-layer graphene at bottom and monolayer $WSe_2$ on top. In addition to protecting $CrI_3$ from degradation under ambient conditions, few-layer graphene acts as a conducting electrode, while monolayer $WSe_2$ provides a strain gauge via measurements of the shift in the exciton energy (under the assumption that $WSe_2$ experiences the same level of strain as $CrI_3$. This 2D heterostructure was first assembled and then transferred on prefabricated circular microtrenches with patterned Au electrodes and Si back gate (Fig. 4c). Fig. 4d shows that the metamagnetic transition field in bilayer $CrI_3$ could be tuned effectively by applying tensile biaxial strain[51].

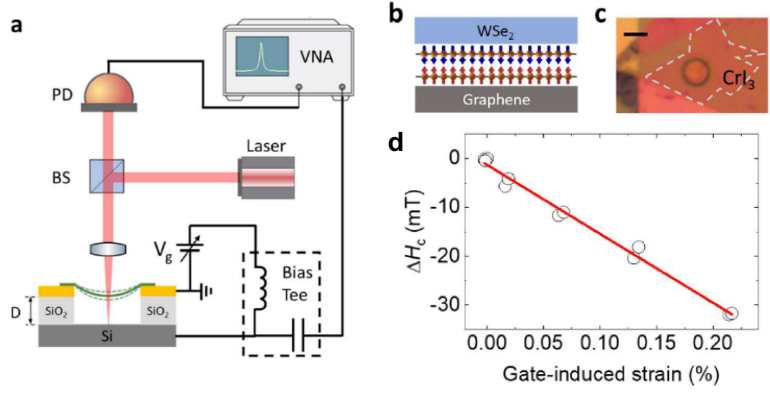

**Figure 4**. Biaxial strain study of CrI$_3$. (a) Schematic of the measurement system. A DC voltage V$_g$ is imposed to apply electrostatic force to the membrane. The laser is to detect both the strain and magnetic ordering. BS: beam splitter; PD: photodetector. (d) The metamagnetic transition field as a function of gate-induced strain (symbols) and the solid line is a linear fit[51].

## 3. THEORY OF MAGNETISM IN 2D

### 3.1. Background

Symmetry breaking in low dimensional systems plays a very special role in condensed matter physics. The spontaneous breaking of a continuous symmetry is not possible in two dimensions at finite temperature, unless long-range interactions come into play. Analogous propositions were posed, in the form of mathematical theorems, in the context of crystalline order by Landau and Peierls, in the context of superconductors and superfluids by Hohenberg, and in the context of magnetism by Mermin and Wagner [8, 52]. The common ground of all these theorems is the existence of gapless collective excitations, the Goldstone modes, each of which is associated with the order parameter of the broken symmetry phase. In the case of magnets, these Goldstone modes are spin waves (or magnons) and in two (and one) dimensions the thermal population of these low energy excitations completely destroys long-range order. This is exemplified by computing the correction to the magnetization within spin wave theory for an isotropic ferromagnet, that yields a divergent result in two dimensions:

$$\delta M(T) \sim \int_0^\infty \frac{k\, dk}{e^{\beta \rho k^2} - 1} \to \infty \qquad (1)$$

where $\delta M(T)$ refers to refers to the correction to the magnetization due to thermal fluctuations, $\rho$ and $\beta$ are the spin wave stiffness and inverse temperature.

***Spin Hamiltonian***. A very wide class of magnetic materials are insulating. Therefore, charge degrees of freedom are frozen and it is possible to describe their magnetic properties in terms of spin Hamiltonians (even in the case of conducting magnetic materials, their magnetic properties can also be described fairly well with effective spin Hamiltonians). So, it is adequate to start our discussion with a brief description of a simplified spin Hamiltonian.

$$H = -\sum_{<ij>} J_{ij}\, \mathbf{S}_i \cdot \mathbf{S}_j - \sum_{<ij>} K_{ij}\, S_i^z S_j^z - \sum_i D_i\, (S_i^z)^2 + \sum_i \mathbf{B} \cdot \mathbf{S}_i \quad (2)$$

The first term describes the exchange (Heisenberg) interactions, the second the anisotropic exchange, the third the single-ion anisotropy, and the fourth introduces the effect of an external magnetic field. We note that additional anisotropic terms could be allowed such as Kitaev [53], biquadratic exchange [54], or Dzyaloshinsky-Moriya [55] interactions (DMI).

The divergent result of Eq. 1 could be avoided if (i) the spin-wave spectra has a gap or (ii) the dispersion of the spin-wave is different. Mechanism (i) corresponds to the existence of a single-ion anisotropy or anisotropic exchange, which yields ferromagnetism in $CrI_3$[3] and $Fe_3GeTe_2$[16]. Mechanism (ii) corresponds to the correction to the spin wave dispersion due to dipolar interactions. Dipolar interactions could allow stabilization of in-plane ferromagnetic order, yet such a scenario has not been confirmed in a 2D van der Waals material to date. With the previous dispersion relation, the correction to the magnetization within the linear spin wave regime becomes at low temperatures $k_B T \ll \Delta$ which yields a finite correction to the maximal magnetization at low enough temperatures, thus allowing for a ferromagnetic state at finite temperatures.

$$\delta M(T) \sim \int_0^\infty \frac{k\, dk}{e^{\beta(\Delta+\rho k^2)}-1} \to \frac{1}{\beta} e^{-\beta \Delta} \quad (3)$$

### 3.2. Origin of magnetic anisotropies

***Single ion anisotropy.*** The simplest anisotropic term that can be written is the so-called uniaxial single ion anisotropy, that takes the form:

$$H_{SIA} = -D \sum_i (S_i^z)^2 \quad (4)$$

The parameter D favors off-plane magnetism for D > 0, whereas it favors in-plane magnetism for D < 0. For D » J, the Heisenberg model reduces to the celeb Ising model. It must be noted the for S = 1/2, $S^2$ = 1/4 and thus the previous term is trivial, yielding that S = 1/2 ferromagnets cannot have single ion anisotropy.

The physical origin of this term is the interplay between the local crystal field δ and the atomic spin-orbit coupling λ. Such anisotropic terms in the Hamiltonian stem from perturbation theory in the high-spin state of the ion, and crucially depend on the spin-orbit coupling of the magnetic ion. We can distinguish between two different cases, systems with orbital degeneracy and without orbital degeneracy. In systems with orbital degeneracy, the single ion anisotropy is first order in λ, yet orbital degeneracy can be easily lifted by a Jahn-Teller mechanism. In the absence of orbital degeneracy, the single ion anisotropy stems (at least) from second order perturbation in λ/δ yielding D ~ (λ/δ)².

Single-ion anisotropy is expected to be strong for transition metals whose crystal field environment has a well-defined symmetry axis as in the 2H-transition metal dichalcogenide structure. In contrast, for approximate octahedral environments such as those in 1T-TMDs or $CrI_3$, the magnitude of the trigonal distortion is expected to substantially impact the value of the single-ion anisotropy *D*.

***Exchange anisotropy.*** The second source of a gap in the spin-wave Hamiltonian is the anisotropic exchange that takes the form:

$$H_{AI} = -K \sum_{<ij>} S_i^z S_j^z \quad (5)$$

where *ij* denotes sum over first neighbors. For a ferromagnet, the previous term favors a parallel off-plane alignment for *K* > 0, whereas for *K* < 0 the system favors in-plane magnetism. We note that this term yields a non-trivial contribution for a *S* = 1/2 system, and thus can yield a magnon gap for a *S* = 1/2 ferromagnet.

Physically, the origin of the anisotropic exchange *K* stems from the connecting atoms between two localized spins. Importantly, in this situation *K* is mainly controlled by the spin orbit coupling of the bonding atom, instead of the magnetic one. A particular example of this is $CrI_3$, where the anisotropy energy is controlled by the strength of the spin-orbit coupling of iodine. Generically, two-dimensional magnets with heavy anions such as Br, I and Te are susceptible to have sizable contributions to the anisotropic exchange due to the large spin-orbit coupling of the ligand anion [56]. The relative strength of the single-ion anisotropy and anisotropic exchange can be estimated from first principles methods, yet their exact values can be sensitive to the details of the method [56, 57].

***Dipolar anisotropy.*** Dipolar interactions are an additional mechanism to stabilize magnetic ordering in two dimensions. In particular, they may allow to stabilize in-plane magnetic ordering at finite temperature. Dipolar interactions favor in-plane arrangement of spins, yielding a Hamiltonian with in-plane rotational symmetry. This leads to the so-called reorientation transition observed in ferromagnetic thin films, that stems from the thermal renormalization of the anisotropy[58], Moreover, dipolar interactions modify the spin-wave spectra so that at low energies the magnon dispersion becomes $E_k \propto k^{1/2}$, yielding the integral in Eq. 1 non divergent. However, van der Waals ferromagnets with in-plane anisotropy and magnetic at zero field are highly elusive, and assessing the existence of in-plane ferromagnetic ordering in some van der Waals systems remains an open question.

### 3.3. Heisenberg Hamiltonian: origin of magnetic exchanges

The strengths and signs of the exchange couplings between different atoms depend on microscopic details, and often arise from a complex interplay between hopping and electronic interactions. Nevertheless, for those cases in the localized limit, i.e. with the active electrons being strongly localized in the magnetic ions, the signs of the different exchange interactions can be predicted using the well-known Goodenough-Kanamori rules[59].

Most of the existing two-dimensional systems that have shown magnetic long-range order in the single-layer limit present structures with some common motifs: hexagonal or triangular lattices in the plane (see Table I), cations in an octahedral environment of their neighboring anions with these octahedra being connected via edge sharing. Most of these systems such as transition metal dihalides and trihalides [60, 61], transition metal dichalcogenides crystallizing in the 1T or 2H structures [62], $Cr_2Ge_2Te_6$ [63], and those of the $AMX_3$ type [64] including phosphosulphides and phosphoselenides can be interpreted in the localized electron limit since most of them are magnetic semiconductors both in their bulk and few-layer form. In this situation, it is important to analyze the possible mechanisms for exchange in such structures. There will be an important contribution coming from direct exchange (metal-metal), and another one coming via an anion, where the cation-anion-cation path forms an angle of approximately 90 degrees. These structural details can be observed in Fig. 5 which depicts the structure of $FePS_3$[65] as an example of a case of a hexagonal in-plane network and the close-up case of two neighboring octahedra sharing an edge.

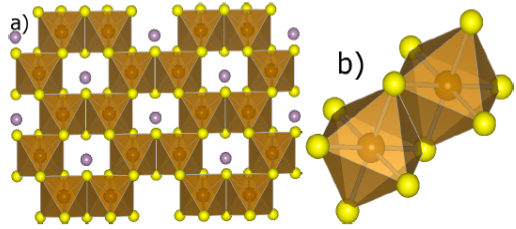

**Figure 5.** a) Structure of FePS$_3$ as seen from the top of the hexagonal plane. b) Two metal atoms surrounded by an ionic octahedral cage. The octahedra are edge sharing. This is the typical coordination in most known 2D vdW magnets. In this situation, competition between metal-metal direct exchange and 90°-superexchange via anions can take place as discussed in the text.

We analyze the situation for various relevant fillings of the external *d* shell of the cations and discuss how ferromagnetic ordering may emerge using halides as a practical example. We stress that it is important to understand the origin of magnetic exchange in these systems in order to enhance the transition temperature, which would require a large FM coupling strength, but also larger moments, and maximized in-plane coordination.

We will limit ourselves to the two-dimensional case using 3*d* electrons as a reference and discuss the evolution with anion size, pressure, etc. We will use mostly transition metal di- and trihalides as an example for the discussion. The important point for following our discussion would be that, as the anion increases in size (going down in their respective column in the periodic table) this gives rise to a larger cation-cation distance, which decreases the strength of the direct exchange. However, the metal-anion-metal interaction is much less affected. Thus, if both interactions have the same sign, increasing the anion size simply leads to a reduction in the magnetic transition temperature. But, if they have opposite signs, as the anion size increases, the sign of the superexchange becomes more important. We will discuss each *d*-filling separately (linking them to materials in which 2D magnetism has been confirmed or theoretically proposed) with Table II compiling all cases analyzed.

| Filling | Cations | Direct exchange | 90° superexchange | Competition |
|---|---|---|---|---|
| $d^3$ | $V^{2+}$, $Cr^{3+}$ | AFM | FM | Yes |
| $d^5$ | $Mn^{2+}$, $Fe^{3+}$ | AFM | FM | Yes |
| $d^6$ | $Fe^{2+}$ | FM | FM | No |
| $d^7$ | $Co^{2+}$ | FM | FM | No |
| $d^8$ | $Ni^{2+}$ | AFM | FM | Yes |

**Table II.** Summary of the magnetic couplings[59] discussed in the text for edge sharing octahedra and various fillings of the 3d shell, with examples of representative cations. AFM stands for antiferromagnetic and FM for ferromagnetic. If the two couplings have opposite signs, competition is active.

***$d^3$ filling.*** There is a competition between an AFM direct exchange and a FM superexchange. Direct exchange decreases its strength as a larger anion is introduced and hence the cation-cation distance is increased. Hence, the tendency for ferromagnetism is enlarged as the unit cell increases. In Cr-trihalides, it is experimentally observed that the FM Curie temperature increases with anion size (17 K for Cl, 33 K for Br, 68 K for I)[66]. In the case of V-dihalides, the AFM Neel temperature decreases as the anion size increases (36 K for Cl, 30 K for Br, 16 K for I)[67], indicating a larger importance of the FM component as the cation-cation distance increases.

*$d^5$ filling.* There is also a competition between AFM direct exchange and FM superexchange. Additionally, there is also a competition in the superexchange between that coming from σ-bonding (mediated by the $e_g$ electrons) which is FM, and that due to π-bonding ($t_{2g}$-mediated) which is AFM. In the case of high spin $d^5$ cations, the FM component becomes important. This competition causes the appearance of a helical phase in $FeCl_3$[68] and a striped phase in Mn-dihalides.

*$d^6$ filling*. In this case there is no competition, both direct exchange and superexchange yield a FM component. That is why in the Fe dihalides, the Curie temperature [57] is larger for $FeCl_2$ (38 K) than for $FeBr_2$ (14 K), a smaller Tc occurs when the cation-cation separation increases. $FePS_3$[2] is another prominent example of FM in-plane ordering with this filling.

*$d^7$ filling.* Again, there is no competition between direct and superexchange, both being FM. An example of this are the Co-dihalides. The Curie temperature is reduced when going from $CoCl_2$ to $CoBr_2$[61, 69] (11 K for $CoI_2$, 19 K for $CoBr_2$ and 25 K for $CoCl_2$) at the same time that the anion size increases leading to a larger Co-Co distance, that decreases the magnetic interaction strength, in particular the direct component.

*$d^8$ filling*. In this case, there is competition between an AFM direct exchange and a FM superexchange. Evidence for this comes from the Ni-dihalides, which are FM in-plane and their Curie temperature increases as the size of the anion does, because the AFM component of the total exchange gets reduced as the cation-cation distance increases. $NiI_2$ has a Curie temperature of 75 K and that of the smaller anion $NiCl_2$ is 52 K[61].

## 4. OVERALL OUTLOOK

### 4.1. Multiferroics

Multiferroics are materials showing a coexisting magnetic and ferroelectric order. Ferroelectric order is the spontaneous development of a finite electric dipole in a material, in analogy with the magnetic ordering of ferromagnet. Ferromagnetism and ferroelectricity are known to obstruct each other. The simplest case is the one of perovskites, where displacive ferroelectricity is favored by an empty d-shell, whereas ferromagnetism requires a partially filled d-shell. As a result, realizing multiferroic orders requires non-displacive mechanism for ferroelectricity, such as charge order, spin-driven, electronic lone pairs or geometric effects. Multiferroic two-dimensional materials would have important applications, including electric reversal of magnetization[70] or electrically controlling an exchange bias[71]. Similar effects have been obtained in two-dimensional heterostructures, as for instance in $CrI_3$ bilayer[39, 40], yet without relying on a multiferroic effect. Multiferroicity has been predicted to intrinsically appear in two-dimensional materials such as transition metal phosphorus chalcogenides[72], $CuBr_2$[73, 74], and $VOI_2$,[75]. Interestingly, artificial multiferroics can be engineered in ferroelectric/ferromagnetic van der Waals heterostructures[72].

### 4.2. Skyrmions

The interplay of ferromagnetic interactions, DMI and an external magnetic field can turn a skyrmion configuration energetically favorable over the spin-spiral and ferromagnetic states.

In addition to their fundamental interest, skyrmions in 2D vdW could provide a new paradigm for low-power data storage. At this point, a few theoretical proposals of skyrmion formation in 2D vdW materials exist. These include twisting in vdW heterostructures, in particular using the example of a ferromagnetic monolayer on top of an antiferromagnetic substrate[76]. An exciting possibility is skyrmion formation via inversion symmetry breaking in Janus monolayers of manganese dichalcogenides that can achieve DMI values comparable to 'traditional' skyrmion-hosting materials [77]. Another proposal shows that, in $CrI_3$ monolayers, skyrmion spin configurations become more stable than FM ones by applying an out-of-plane electric field [78].

Recently, the first experimental observation of magnetic skyrmions in the 2D vdW ferromagnet $Fe_3GeTe_2$ was reported using high-resolution scanning transmission X-ray microscopy (STXM) and Lorentz transmission electron microscopy measurements (Fig. 6a). A skyrmion crystal state can be generated both dynamically using current pulses and statically using canted magnetic fields (Fig. 6b)[79].

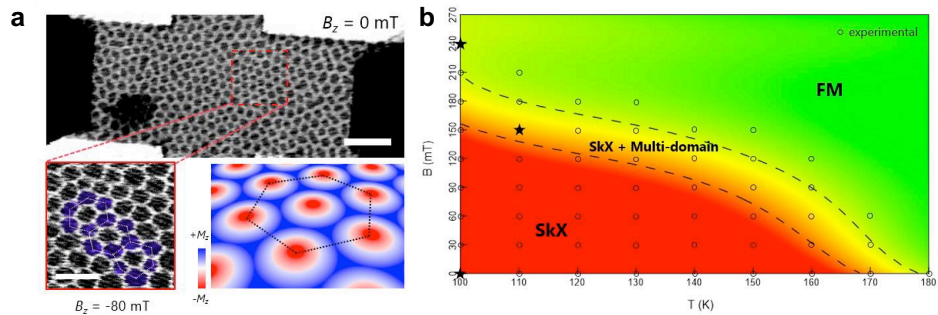

**Figure 6.** Magnetic skyrmion lattice phase in FGT. a) Representative STXM image of skyrmion lattice stabilized over the whole FGT at $B_z$=0 mT and T=100 K. Scale bar, 2 µm. b) Experimental phase diagram of magnetic configurations as a function of temperature and magnetic field[79].

### 4.3 Quantum spin liquids

Quantum spin liquids (QSLs) are a class of quantum disordered phases where reduced dimensionality, geometric frustration and quantum fluctuations completely destroy long range magnetic order down to zero temperature. QSLs are intriguing as they exhibit topological entanglement entropy as well as fractionalized excitations that obey emergent gauge fields (see Ref. [80] for a recent review). Experimental search for QLSs mostly targeted layered magnets whereas the majority of the theoretical studies are for two-dimensional models. Therefore, 2D vdW magnets provide a unique opportunity to discover new QSLs. Two promising routes are the (i) honeycomb lattices with Kitaev exchange and (ii) triangular lattices with frustrated interactions. Recently, a strongly spin–orbit-coupled vdW Mott insulator, $\alpha$-$RuCl_3$, has emerged as a prime candidate for hosting an approximate Kitaev QSL[81-85]. Figure 7a shows the thermal hall conductivity when applying tilted magnetic field on α-$RuCl_3$ at different temperatures and the half-integer plateau indicates the Majorana fermion, which is a sign of Kitaev spin liquid phase. Figure 7b shows the phase diagram of $\alpha$-$RuCl_3$ in a field tilted at θ = 60° (right inset). Below T $\approx$ $J_K/k_B$ $\approx$ 80 K, the spin-liquid (Kitaev paramagnetic) state appears and the half-integer quantized plateau of the 2D thermal Hall conductance is observed in the red area.

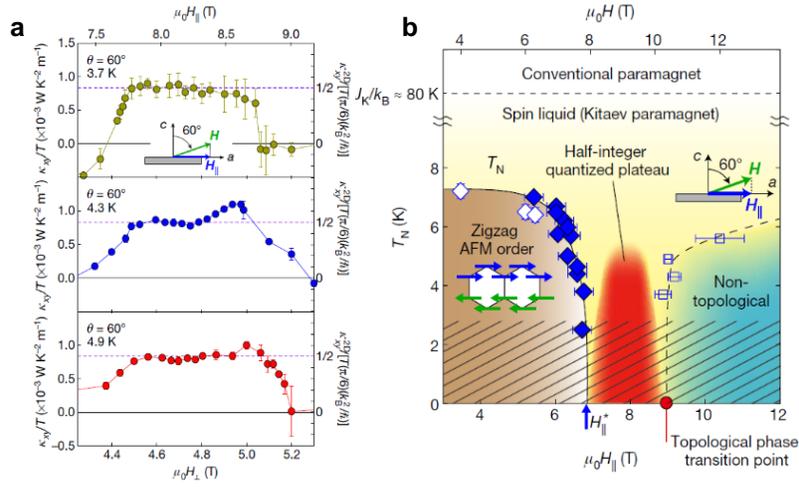

**Figure 7**. Half-integer thermal quantum Hall Effect in $\alpha$-$RuCl_3$ (a) Half-integer thermal hall conductivity indicates the Majorana fermion, which is a sign of spin liquid phase. (b) Phase diagram of $\alpha$-$RuCl_3$ in a field tilted at θ = 60°. Below T ≈ $J_K/k_B$ ≈ 80 K, the spin-liquid (Kitaev paramagnetic) state appears[83].

### 4.4. From the synthesis perspective

As mentioned above, the current technologies only enable the field to produce vdW crystals of the magnetic materials described above. While this is a logical choice for fundamental research in order to identify the most promising 2D magnetic materials, more comprehensive crystal growth studies are needed to understand how crystal growth techniques, thermal profile, and precursor types ultimately influence the fundamental behavior of vdW magnetic crystals. Current literature heavily relies on half a century-old crystal growth methods. While these techniques are well-established to produce these crystals, prior literature has given very little attention to magnetic quantum phenomena in 2D. Thus, more careful crystal growth studies are needed to pinpoint how defect density can be reduced, how crystalline quality can be improved, and magnetic impurities eliminated. Clearly, new crystal growth techniques or recipes will be required to produce recently predicted magnetic crystals. This is a challenging task especially for ternary and quaternary systems wherein many different phases or compositions might energetically compete with each other to produce mixed-phase crystals.

In the very big picture, these crystal growth methods and isolation of mono- and few-layers of 2D magnets by exfoliation techniques present added complexities in translating these fundamental results from the laboratory setting to applications and eventually technology development. To this end, large scale growth methods will be required, in order to produce them at wafer scales. This is a big ask from the materials synthesis community when the number of theoretically predicted 2D magnetic crystals is still increasing on a daily basis. As such, fast progress is needed to quickly identify the champion magnetic materials and develop more focused synthesis techniques to produce them at large scales (centimeter to 2 inches wafer). Here, the grand challenge will likely be in retaining their structural quality and defect profiles while increasing their lateral sizes to wafer scales. Nevertheless, general 2D growth techniques specific to halide, phosphosulfide, or tellurium based 2D magnetic material systems will greatly benefit the 2D magnetism community in the long run by offering the foundations of 2D growth in these material systems.

Still many questions emerge in the synthesis of atomically thin large area 2D magnetic layers; Can large area synthesis produce 2D sheets with environmental stability properties comparable to those in bulk crystals? Can we eliminate large defect densities like those observed in large area 2D transition metal dichalcogenide systems? Can we engineer defects, strain, or pressure in these 2D magnets during synthesis by using a different choice of substrates, growth cooling profiles, or introduced defects? Can these sheets be synthesized on arbitrary substrates? Can we alloy 2D magnetic materials to unleash exciting opportunities similar to those realized in traditional materials alloying? These and many other overwhelming but equally exciting questions are awaiting the materials synthesis community and only brilliant work by researchers in the field will be capable to provide solid answers.

## 5. ACKNOWLEDGMENTS

S.T acknowledges support from DOE, NSF DMR 1552220, DMR 1955889, DMR 1904716, and NSF CMMI 1933214. D.D. acknowledges ASU for startup funds. A.B.S and O.E. acknowledge support from NSF 1904716. R.C. acknowledges support from the Alfred P. Sloan Foundation. RC and QS work was supported by the STC Center for Integrated Quantum Materials, NSF DMR 1231319. V.P. acknowledges support from the MINECO of Spain through the project PGC2018-101334-B-C21. J.L.L. acknowledges support from the Aalto Science-IT project.

## 6. DATA AVAILABILITY STATEMENT.

Data sharing is not applicable to this article as no new data were created or analyzed in this study.